\documentclass[journal]{IEEEtran}

\usepackage{xcolor,soul,framed} %,caption

\colorlet{shadecolor}{yellow}
\usepackage[pdftex]{graphicx}
\graphicspath{{../pdf/}{../jpeg/}}
\DeclareGraphicsExtensions{.pdf,.jpeg,.png}

\usepackage[cmex10]{amsmath}
\usepackage{array}
\usepackage{mdwmath}
\usepackage{mdwtab}
\usepackage{eqparbox}
\usepackage{url}

\usepackage{multirow}
\usepackage{graphicx}
\usepackage{amsmath}
\usepackage{booktabs}
\usepackage{adjustbox}

\begin{document}
% \bstctlcite{IEEEexample:BSTcontrol}
    \title{Bimodal Connection Attention Fusion for Speech Emotion Recognition}  
      \author{Jiachen Luo,
      Huy Phan,
      Lin Wang,
      Joshua D. Reiss*

% \thanks{Manuscript submitted Dec , 2024.}
  
%   \thanks{M. Roberg is with TriQuint Semiconductor, 500 West Renner Road Richardson, TX 75080 USA (e-mail: michael.roberg@tqs.com).}% <-this % stops a space
%   \thanks{T. Reveyrand is with the XLIM Laboratory, UMR 7252, University of Limoges, 87060 Limoges, France (e-mail: tibault.reveyrand@xlim.fr).}%
%   \thanks{I. Ramos and Z. Popovic are with the Department of Electrical, Computer and Energy Engineering, University of Colorado, Boulder, CO, 80309-0425 USA (e-mail: ignacio.ramos@colorado.edu; zoya.popovic@colorado.edu).}% <-this % stops a space
%   \thanks{E. Falkenstein is with Qualcomm Inc., 6150 Lookout Road
% Boulder, CO 80301 USA (e-mail: erez.falkenstein@gmail.com).}

\thanks{The work does not relate to Huy Phan’s position at Meta, Paris, France, e-mail: huyphan@meta.com. \\
Jiachen Luo, Lin Wang and Joshua Reiss are with the Centre for Digital Music, Queen Mary University of London, London, United Kingdom, UK, e-mail: \{jiachen.luo, lin.wang, joshua.reiss\}@qmul.ac.uk. \\
* corresponding author
}
}

% \markboth{IEEE TRANSACTIONS ON Multimedia, February~2025}{Roberg \MakeLowercase{\textit{et al.}}: High-Efficiency Diode and Transistor Rectifiers}

\maketitle

% === ABSTRACT ====================================================================
\begin{abstract}
Multi-modal emotion recognition is challenging due to the difficulty of extracting features that capture subtle emotional differences. Understanding multi-modal interactions and connections is key to building effective bimodal speech emotion recognition systems. In this work, we propose Bimodal Connection Attention Fusion (BCAF) method, which includes three main modules: the interactive connection network, the bimodal attention network, and the correlative attention network. The interactive connection network uses an encoder-decoder architecture to model modality connections between audio and text while leveraging modality-specific features. The bimodal attention network enhances semantic complementation and exploits intra- and inter-modal interactions. The correlative attention network reduces cross-modal noise and captures correlations between audio and text. Experiments on the MELD and IEMOCAP datasets demonstrate that the proposed BCAF method outperforms existing state-of-the-art baselines.
\end{abstract}


\IEEEpeerreviewmaketitle

\section{Introduction}
\begin{figure*}
\centering
\includegraphics[width=0.8\linewidth]{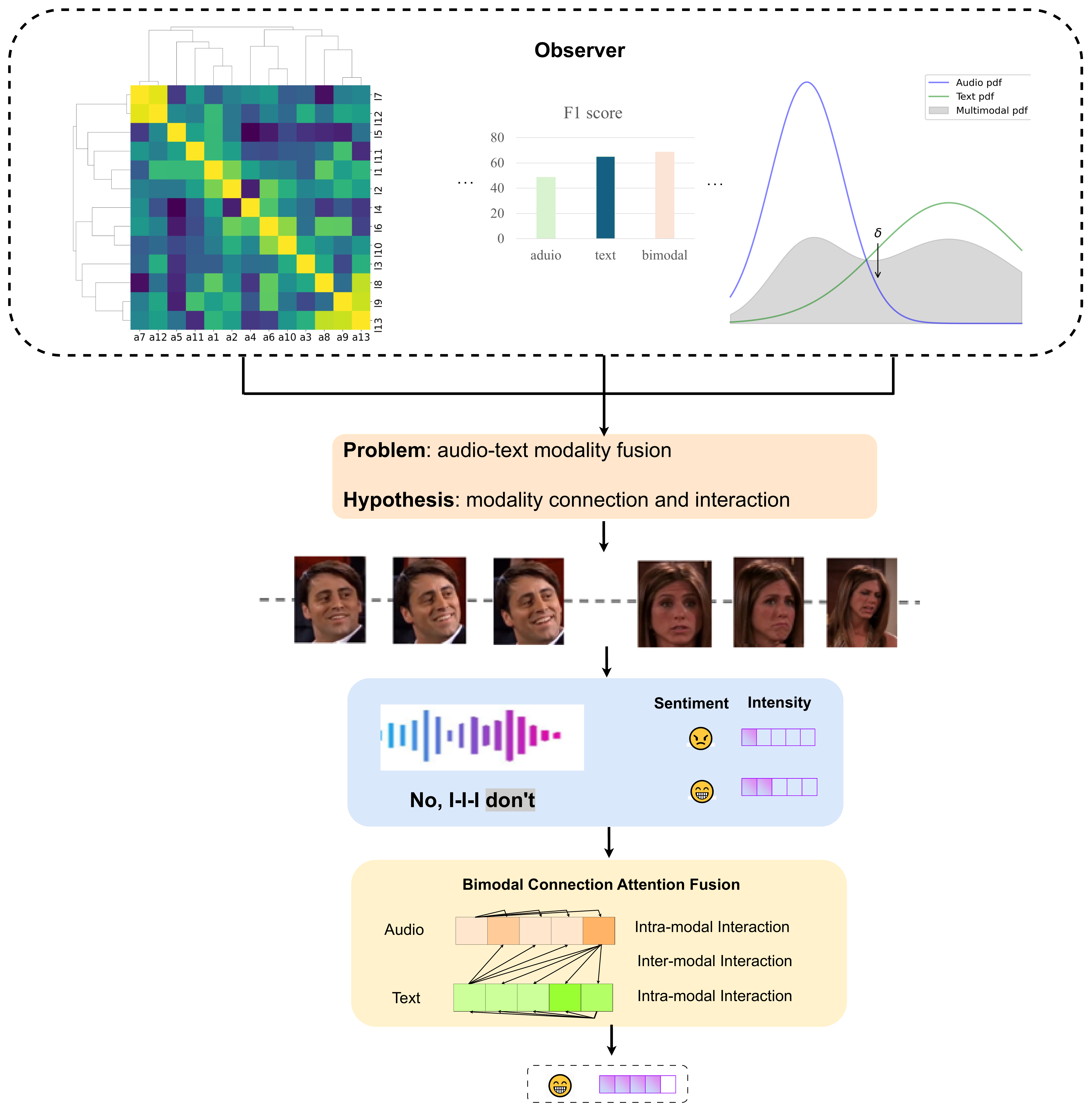}
\caption{Architecture of the proposed Bimodal Connection Attention Fusion (BCAF) method. The method consists of three modules: the unimodal representation module, the connection attention fusion module, and the classification module. The core connection attention fusion module includes the interactive connection network, the bimodal attention network, and the correlative attention network, with details depicted in Figs.~2, 3, 4 and 5. The unimodal audio representation $H_a$ and text representation $H_l$ are input into the interactive connection network, the correlative attention network, and the bimodal attention network.}
\end{figure*}
\IEEEPARstart{T}{he} proliferation of mobile internet and smartphones has led to the widespread use of social networking platforms where users create and share content in various modalities, such as audio, text, and video. Extracting and analyzing emotions from this multimodal content has extensive applications in human-computer interaction, surveillance, robotics, and gaming \cite{cai2023emotion, singh2022systematic, krishnan2025linecongraphs}. However, effectively integrating multiple modalities remains a significant challenge in emotion recognition research.

Previous multi-modal emotion recognition methods have achieved good performance \cite{singh2023universality}; however, key challenges remain in multi-modal emotion recognition. Different modalities require independent preprocessing and feature extraction designs due to their heterogeneous nature \cite{pan2023review, zhao2023tdfnet, tellamekala2023cold}. To develop a model that is both applicable and generalizable across individual modalities and fusion models, it is essential to learn modality interactions and connections for learning discriminative emotional content. 

In multi-modal learning, modality connection in multi-modal learning refers to the extent to which information is shared across modalities, shaping a unified representation \cite{baltruvsaitis2018multimodal}. Unlike correlation, which quantifies modality dependence, modality connection captures the semantic alignment between modalities. For example, a higher vocal pitch may correlate with excitement, but modality connection determines if the audio tone, facial expressions, and text consistently reinforce that excitement. A strong modality connection ensures meaningful interaction among modalities, whereas a weak connection may lead to ambiguity. Leveraging modality connections enables a more nuanced understanding of emotions, improving the robustness and accuracy of emotion recognition systems.

% While correlation measures numerical relationships, modality connection evaluates how different modalities provide reinforcing signals. 

%For instance, in detecting sadness, shared signals across modalities—such as a soft, low-pitched voice and distressing text phrases—strengthen emotional interpretation. A high modality connection suggests alignment between modalities, enhancing recognition accuracy. Conversely, if a text expresses distress while the tone remains neutral, the weak connection could cause misinterpretation. Leveraging modality connections enables a more nuanced understanding of emotions, improving the robustness and accuracy of emotion recognition systems.

On the other hand, modality interaction refers to the relationships and dependencies between modalities, which can vary from strong correlations to weak or even conflicting signals \cite{rasenberg2020alignment}. While modality connection emphasizes shared content, modality interaction captures how different modalities influence one another. For example, imagine a meeting scenario where a participant says, ``That’s great'' in a flat, monotone voice while avoiding eye contact and crossing their arms. The verbal content suggests positivity, but the tone and body language convey disengagement or sarcasm. Such conflicting signals highlight the complexity of multi-modal interactions. Accurately interpreting these interactions requires understanding the context and modeling how modalities influence one another.

Two core types of interactions are frequently encountered in multi-modal learning: intra-modal and inter-modal interactions \cite{zhou2023intra,lin2022modeling}. Intra-modal interactions measure relationships within the same modality, while inter-modal interactions capture relationships between different modalities. Understanding and learning these interactions are essential for building an effective multi-modal emotion recognition system that achieves three key goals: multi-modal integration, robustness, and contextual awareness.

Many multi-modal emotion recognition simply focused on exploring interactions across different modalities while often overlooking the importance of modality connections. Existing approaches simply concatenated data for joint fusion and aggregates decisions from various modalities through weighted averaging \cite{pan2023review, ezzameli2023emotion, luo2023cross}, neglecting intra-modal interactions. Recently attention mechanisms have gained attraction due to their effectiveness in modeling modality interactions~\cite{chudasama2022m2fnet,fan2025coordination,luo2025triagedmsa}. However, such approaches often lack a deeper understanding of modality interactions and connections. This limitation can lead to several issues, including suboptimal connection strategies, loss of critical cross-modal information, and reduced robustness when handling conflicting or incomplete data from specific modalities. Addressing these challenges is crucial for enhancing the performance and generalizability of multi-modal emotion recognition systems in real-world applications. 

%For instance, Chudasama et al. proposed the multimodal fusion network, which extracts emotion-relevant features from visual, audio, and text modalities, leveraging the multi-head attention-based fusion mechanism to capture both intra- and inter-modal interactions \cite{chudasama2022m2fnet}. 

% We propose the connection attention fusion module to learn modality connections, filter noise from cross-modal relationships, and effectively model both intra- and inter-modal interactionsbetween audio and text.
 
Motivated by the above observations (Fig. ~1), we propose a Bimodal Connection Attention Fusion (BCAF) framework for bimodal emotion recognition, which consists of three key modules: the uni-modal representation module, the connection attention fusion module, and the classification module (Fig.~2). The contribution is summarized as follows. First, we propose the interactive connection network to capture modality-specific features and analyze multi-modal connection between audio and text. Second, We investigate the bimodal attention network, which assigns dynamic weights to comprehensively learn intra- and inter-modal interactions between audio and text. Finally, We design the correlative attention network to effectively filter out incorrect cross-modal relationships and enhance learning of both intra- and inter-modal interactions between audio and text. Experimental results validate the effectiveness of the proposed method on public datasets. 

%First, we use the interactive connection network to learn modality-specific features and analyze modality connections between audio and text. Second, we introduce self- and cross-attention mechanisms to learn intra- and inter-modal interactions between audio and text. Finally, we employ the correlative attention network to filter out incorrect cross-modal relationship and facilitate intra- and inter-modal interactions between audio and text. The main contributions of this paper are summarized as follows:

%\begin{itemize}
%    \item We propose the interactive connection network to capture modality-specific features and analyze multi-modal connection between audio and text.
    
%    \item We investigate the bimodal attention network, which assigns dynamic weights to comprehensively learn intra- and inter-modal interactions between audio and text.
    
%    \item We design the correlative attention network to effectively filter out incorrect cross-modal relationships and enhance learning of both intra- and inter-modal interactions between audio and text.
%\end{itemize}

The remainder of this paper is organized as follows: Section II presents a brief literature review. Section III describes our method in detail. Section IV outlines the experiments conducted. Section V discusses the results. Finally, Section VI provides conclusions based on this work.

\section {Related Work}
Emotion recognition in conversations has found widespread applications across various fields \cite{singh2023universality}. Humans convey emotions through multiple modalities, including speech, facial expressions, and body postures \cite{ezzameli2023emotion, zhu2022multimodal, zhang2023learning}. Among these, speech is a crucial modality, carrying emotional cues through both paralinguistic features and linguistic content. Since different modalities provide complementary information, relying on a single modality is often insufficient for accurate emotion recognition \cite{pan2023review, jabeen2023review}. Therefore, combining information from multiple modalities enhances the ability to discern emotions, as each modality can complement or augment others, providing richer emotion-relevant information. Consequently, multi-modal approaches generally yield superior results compared to uni-modal methods, leading to substantial efforts in developing and exploring multi-modal fusion techniques for more accurate emotion recognition in conversations.

Multi-modal fusion methods are broadly categorized into early fusion, late fusion, and hybrid fusion \cite{cai2023emotion, singh2022systematic, singh2023universality, li2024fer}. Early fusion combines features from different modalities at the input level but often overlooks complex inter-modal dependencies \cite{poria2018meld, gunes2005affect}. Late fusion integrates decision-level outputs from unimodal classifiers \cite{busso2004analysis, jin2015speech}, simplifying the process but limiting cross-modal interactions. Hybrid fusion addresses these limitations by combining intermediate representations of multiple modalities \cite{zhang2017learning, wollmer2013lstm}.

Recent multi-modal emotion recognition approaches employ advanced fusion strategies to capture intra- and inter-modal interactions. Transformer-based models \cite{li2023transformer} and gated recurrent units \cite{hazarika2018icon} have been explored for cross-modal fusion. Multi-head attention mechanisms, as used in M2FNet \cite{chudasama2022m2fnet}, effectively learn intra- and inter-modal relationships. Similarly, HCAM \cite{dutta2023hcam} leverages recurrent and co-attention networks for improved fusion. CFN-ESA \cite{li2024cfn} introduces emotion-shift awareness in cross-modal fusion, while TelME \cite{yun2024telme} enhances non-verbal modalities through knowledge transfer. Mamba \cite{shou2024revisiting} employs probability-guided multi-modal fusion to maintain consistency across modalities, and AGF-IB \cite{shou2024adversarial} uses contrastive learning for capturing inter-class and intra-class semantic relationships.

Despite these advances, challenges remain. Many methods underutilize audio representations, often prioritizing text due to its rich contextual information. Additionally, existing approaches struggle to fully capture cross-modal dependencies across different levels, limiting their ability to model long-term emotional context. To address these issues, we propose the BCAF method, which enhances modality interactions for bimodal speech emotion recognition.

\section{Methodology}
% Our goal is to infer the emotions of utterances presented in multi-turn, multi-speaker conversations. The emotion recognition in conversations task includes \( C \) emotion categories, which we denote as the set \( E = \{ E_1, E_2, \ldots, E_C \} \). We define a dialogue with \( N \) utterances \( u^1_m, u^2_m, \ldots, u^N_m \), where \( m \in \{a, l\} \) represents audio and textual modalities, respectively. For each utterance \( u^n_m \), \( 1 \le n \le N \), an emotion label \( E_c \), \( 1 \le c \le C \), is assigned.

Our proposed BCAF method simultaneously models modality connections and interactions for bimodal speech recognition system. Fig. 1 illustrates the architecture of BCAF, which consists of the uni-modal representation module, the connection attention fusion module, and the classification module. The core connection attention fusion module comprises of the the interactive connection network, the bimodal attention network and the correlative attention network (see Figs.~3-5). 

%Detailed explanations of BCAF are provided in the following subsections. 

% The audio and text modalities are first processed through separate encoders to extract uni-modal representations, which are then fused within the connection attention fusion module to predict emotions. 

\subsection{Uni-modal Representation Module}
\subsubsection{Acoustic Encoder}
We use the large wav2vec model as the audio modality encoder to obtain a 1024-dimensional utterance-level audio representation from raw audio signals \cite{baevski2020wav2vec}. The large wav2vec model is an advanced self-supervised learning framework designed for speech representation learning. It consists of three key components: a feature encoder, a Transformer-based contextual representation module, and a quantization module. In total, 1024-dimensional utterance-level acoustic features were extracted ($ H_a $). 

 % We employthe large wav2vec model as the acoustic feature extractor to 1024-dimensional utterance-level representations ($a^{wav2vec}$) from raw audio signal.
 
\subsubsection{Textual Encoder}
We use the RoBERTa model as the text modality encoder to extract a 1024-dimensional utterance-level text representation from raw text \cite{liu2019roberta}. RoBERTa takes the utterance transcript as input and generates rich contextual representations from the final four layers. This process produces four 1024-dimensional vectors for each token in the input. We then average these vectors to obtain a contextual utterance feature vector with a dimension of 1024 (\( H_l \)).

\begin{figure*}
\centering
\includegraphics[width=1\linewidth]{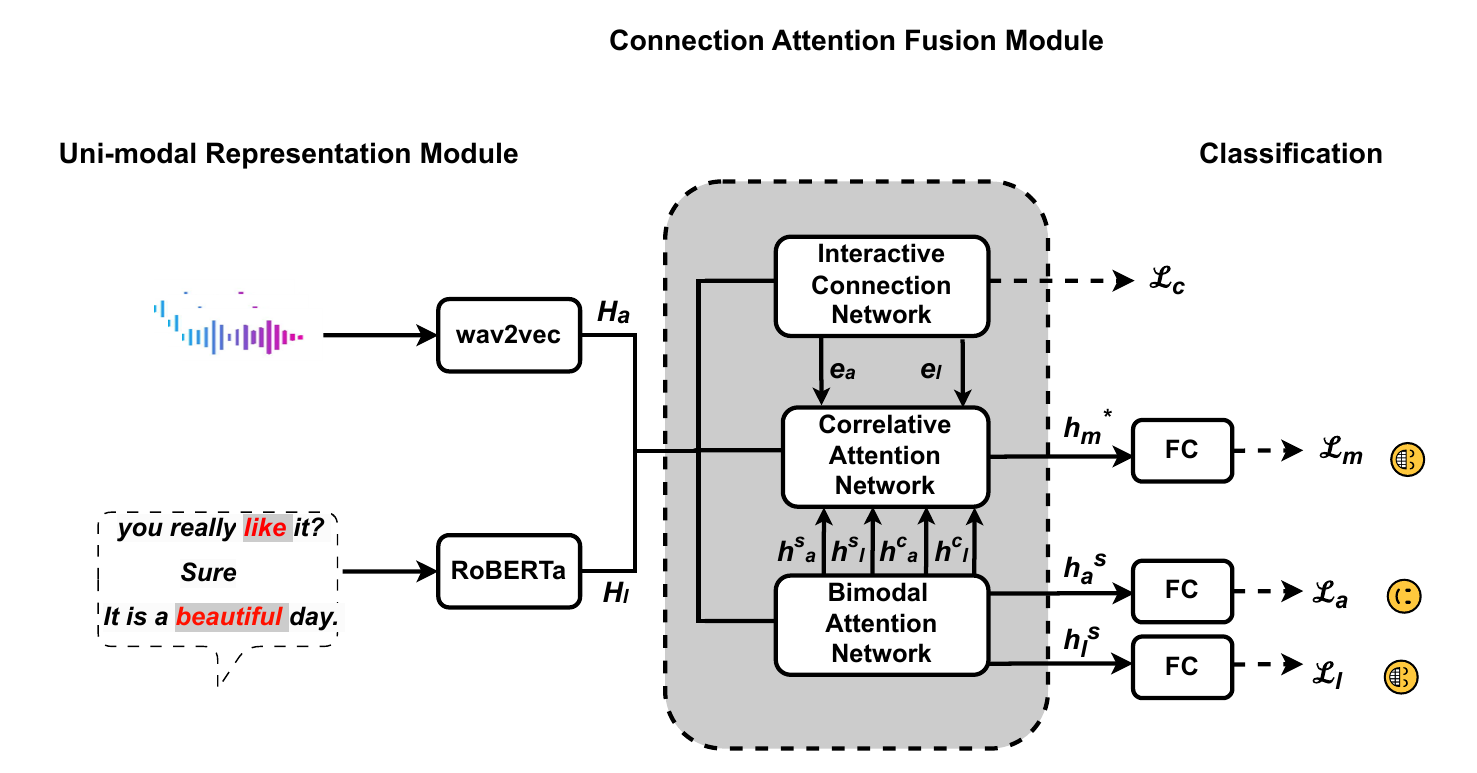}
\caption{Architecture of our proposed the Bimodal Connection Attention Fusion method. The method consists of three modules, the uni-modal representation, the connection attention fusion module and the classification module. The uni-modal audio representation $H_a$ and text representation $H_l$ are inputted into all the interactive connection network, correlative attention network and bimodal attention network. The core connection attention fusion module includes the interactive connection network, the bimodal attention network and  correlative attention network, with the details depicted in Figs. 3-5, respectively.}
\label{fig:BCAF}
\end{figure*}

\subsection{Connection Attention Fusion Module}
We propose the connection attention fusion module to learn modality connections and interactions between audio and text for bimodal speech emotion recognition. The module comprises three main components: the interactive connection network, the bimodal attention network, and the correlative attention network (see Figs.~3-5). Detailed explanations of these components are provided in the following subsections.

\subsubsection{Interactive Connection Network}
The interactive connection network uses an encoder-decoder architecture to learn modality connections between audio and text (see Fig.~3). Both the encoder and decoder consist of three fully connected layers, with each layer applying a linear transformation followed by a non-linear ReLU activation function. In the encoder, the fully connected layers progressively reduce the dimensionality of the input uni-modal representation \( H_m \), extracting important features and compressing them into a fixed-size latent vector \( e_m \). This process of hierarchical dimensionality reduction enables the encoder to capture the essential characteristics of the input modality while filtering out irrelevant or redundant information.

The decoder, which also consists of three fully connected layers, reverses this process by gradually increasing the dimensionality of the latent vector \( e_m \). It reconstructs the original uni-modal representation \( d_m \) using the latent features as a guide. Each layer in the decoder applies a linear transformation and ReLU activation function to ensure that the reconstructed output closely resembles the original input. This encoder-decoder architecture effectively enables feature compression and reconstruction, facilitating the robust learning of modality connections.
\begin{figure}[!ht]
    \centering
    \includegraphics[width=1\linewidth, height=0.35\textheight]{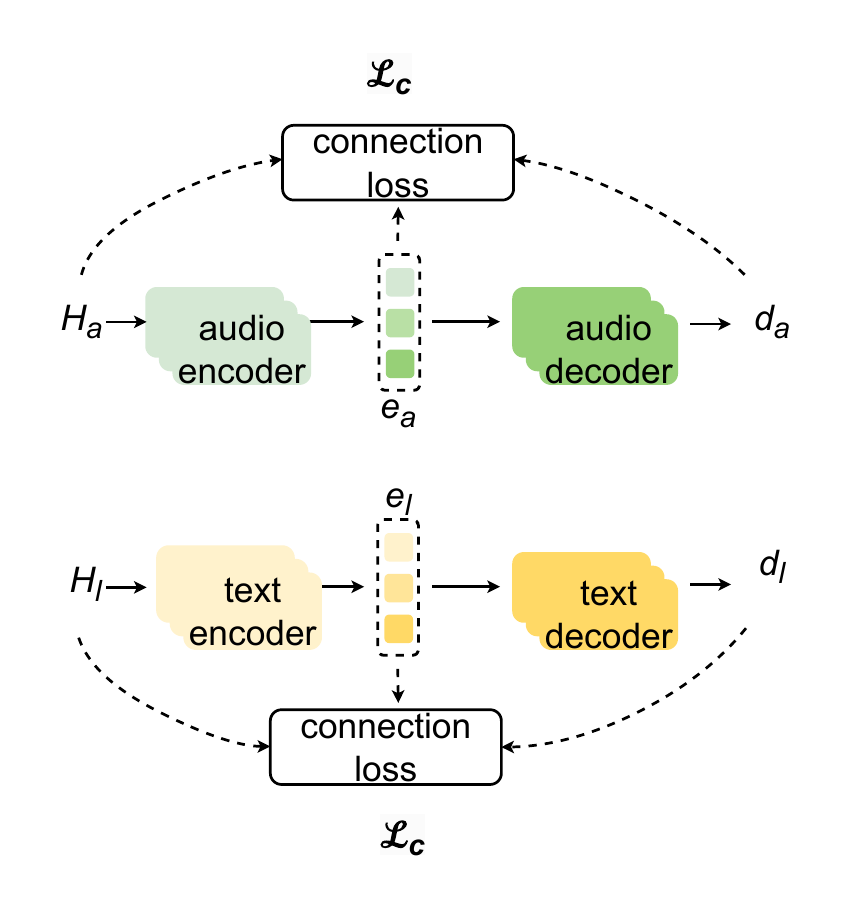}
    \caption{Update scheme of the interactive connection network.}
    \label{fig:interaction_network}
\end{figure}

This symmetric encoder-decoder architecture facilitates efficient feature extraction, compression, and reconstruction. By designing the encoder to focus on compact and meaningful representations and the decoder to restore the input features accurately, the network ensures that modality-specific information is preserved while enabling cross-modality learning.

The interactive connection network adopts an architecture comprising stacked fully connected layers followed by a dropout layer in both the encoder and decoder (see Fig. 3). For each modality \( m \in \{ a, l \} \), a simple encoder-decoder is formulated as:
\begin{equation} 
e_m = E_m(H_m, \theta_m^e)
\end{equation} 

\begin{equation} 
d_m = D_m(e_m, \theta_m^d)
\end{equation}
where \( E_{m}(\cdot) \) denotes the encoder function for modality \( m \), with \( \theta_m^e \) as its trainable parameters, and \( D_{m}(\cdot) \) represents the decoder function for modality \( m \), with \( \theta_m^d \) as its trainable parameters. The dimensions of \( H_m \) and \( d_m \) are both 1024. The dimension of \( e_m \) is 512.

The objective function models the learning problem using the self-supervised spirit. We design the connection loss function to maximize modality connections between audio and text. It is defined as follows:

% \begin{equation} 
% \mathcal{L}_c = 
% \begin{aligned}
% &\bigg( \| H_a - d_a \|_F^2 + \| H_l - d_l \|_F^2 \\
% &+ \mu \bigg( \| I_e - c_a^\top e_l \|_F^2 \\
% &\quad - \| I_d - d_a^\top d_l \|_F^2 \bigg) \bigg)
% \end{aligned}
% \end{equation}

{\scriptsize
\begin{equation} 
\mathcal{L}_c = \left( \| H_a - d_a \|_F^2 + \| H_l - d_l \|_F^2 +\\
\mu \left( \| I_e - e_a^\top e_l \|_F^2 - \| I_d - d_a^\top d_l \|_F^2 \right) \right)
\end{equation}}where \( \| \cdot \|_F^2 \) is the squared Frobenius norm, which calculates the sum of squared elements in a matrix. \( I_e \) and \( I_l \) represent the identity matrix, and \( \mu \) is a non-negative hyperparameter that controls the balance between terms. 

The first term of the objective function, \( \| H_a - d_a \|_F^2 + \| H_l - d_l \|_F^2 \), measures the reconstruction error between the original input representations and their reconstructed outputs for both the audio (\( H_a, d_a \)) and text (\( H_l, d_l \)) modalities. By minimizing this term, the model ensures that the decoder can accurately reconstruct the original input representations, preserving the modality-specific information unique to each modality. This encourages the encoder-decoder mechanism to retain the essential features of each modality while discarding irrelevant or redundant information. As a result, the model achieves high-quality reconstruction for both modalities, which is critical for ensuring robust performance in downstream tasks.

The second term, \( \| I_e - e_a^\top e_l \|_F^2 - \| I_d - d_a^\top d_l \|_F^2 \), focuses on learning meaningful connectionsbetween audio and text. The first component, \( \| I_e - e_a^\top e_l \|_F^2 \), aligns the latent representations (\( e_a, e_l \)) in the feature space, ensuring strong modality connections during the encoding process. The second component, \( \| I_d - d_a^\top d_l \|_F^2 \), maintains these connections in the reconstructed space by aligning the outputs (\( d_a, d_l \)). Together, these terms promote consistency between the latent and reconstructed spaces, enabling the model to capture robust cross-modal relationships. This ensures that the learned modality connections are both meaningful and preserved across all stages of the network.

\subsubsection{Bimodal Attention Network}
The bimodal attention network introduces self-attention and cross-attention mechanisms to learn intra- and inter-modal interactions between audio and text. Fig.~5 illustrates the architecture of the bimodal attention network. The input to this network consists of queries, keys, and values. The dot product of the query and each key is computed, and a softmax function is applied to generate weights for the values \cite{vaswani2017attention}. The bimodal attention network comprises stacked self-attention and cross-attention layers, along with feed-forward layers. 

The core idea of this module is to learn intra- and inter-modal interactions between audio and text, then  propagate information from both modality-specific patterns and modality associations based on the attention weights. Technically, the self-attention layer aims to learn intra-modal interactions within each modality, such as audio or text. The query, key, and value are derived from the same modality. Given weight matrices \( W^Q_m \), \( W^K_m \), and \( W^V_m \), the modality representation \( H_m \) is projected into the query matrix (\( Q_m \)), key matrix (\( K_m \)), and value matrix (\( V_m \)) through linear projections without bias. The self-attention representation can then be summarized as follows:
\begin{equation}
\zeta H_m = \text{softmax} \left( \frac{Q_m K_m^T}{\sqrt{d_k}} \right) V_m
\end{equation}
where \( \zeta H_m \) represents the self-attention representation with a dimensionality of 1024. $d_k$
represents the dimension of the key vector.

\begin{figure*}
    \centering
    \includegraphics[width=1\linewidth, height=0.35\textheight]{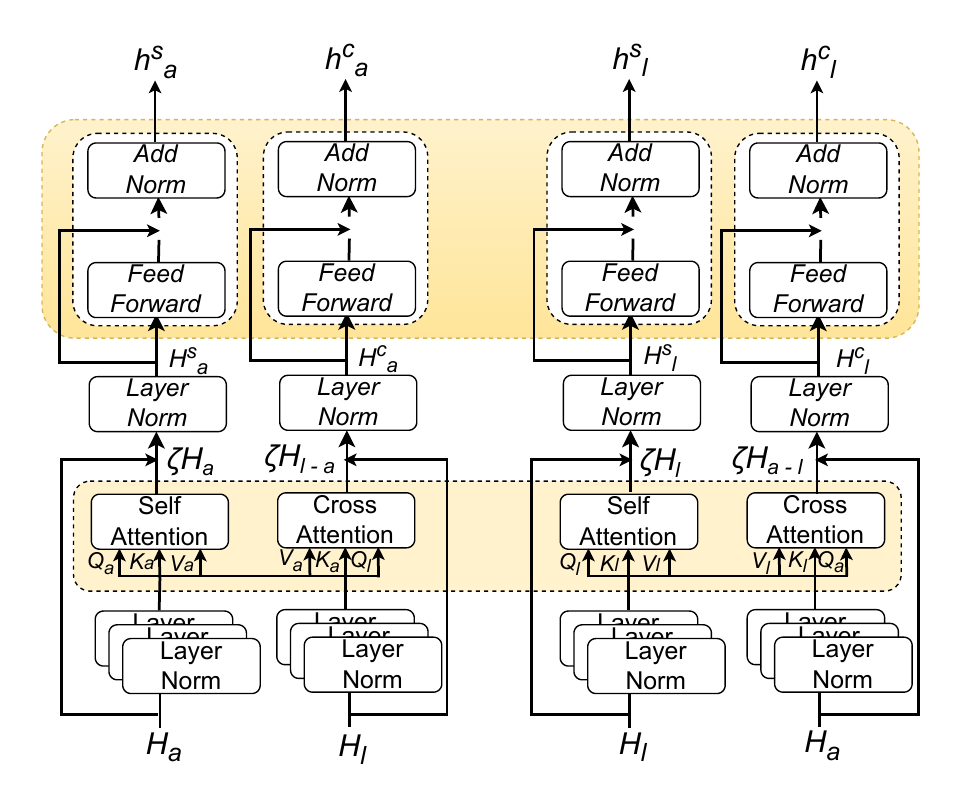}
    \caption{Update scheme of the bimodal attention network.}
    \label{fig:bimodal_attention}
\end{figure*}

To further enhance the representation capacity, the self-attention representation is passed through a LayerNorm layer followed by an AddNorm layer to obtain the enhanced self-attention representation.

\begin{equation} 
H^{s}_m = \text{LayerNorm}(H_m \oplus \zeta H_m) 
\end{equation} 

\begin{equation} 
h^s_{m} = \text{AddNorm}(H^{s}_m \oplus \text{FeedForward}(H^{s}_m))
\end{equation} 
where \( h^s_{a} \) and \( h^s_{l} \) represent the enhanced self-attention representations for audio and text, respectively. The dimensions of $h^s_{a}$ and $h^s_{l}$ are both 1024.

Parallel to the self-attention layer, the cross-attention layer captures inter-modal interactions between audio and text. It learns associations between the two modalities and propagates information from one modality to the other based on these associations. Specifically, the cross-attention mechanism follows a similar principle to the self-attention mechanism, with the key difference being that the query, key, and value are derived from different modalities. The enhanced cross-attention representation is summarized as follows:

% \begin{equation} 
% \zeta H_{l-a} = \text{softmax}(Q_l K^\mathsf{T}_a/\sqrt{d})V_a
% \end{equation}

\begin{equation}
\zeta H_{l-a} = \text{softmax} \left( \frac{Q_l K^\mathsf{T}_a}{\sqrt{d_k}} \right) V_a
\end{equation}

\begin{equation} 
H^{c}_a = \text{LayerNorm}(H_a \oplus \zeta H_{l-a}) 
\end{equation} 

\begin{equation} 
h^c_{a} = \text{AddNorm}(H^{c}_a \oplus \text{FeedForward}(H^{c}_a))
\end{equation} 

% \begin{equation} 
% \zeta H_{a-l} = \text{softmax}(Q_a K^\mathsf{T}_l/\sqrt{d})V_l
% \end{equation}

\begin{equation}
\zeta H_{a-l} = \text{softmax} \left( \frac{Q_a K^\mathsf{T}_l}{\sqrt{d_k}} \right) V_l
\end{equation}

\begin{equation} 
H^{c}_l = \text{LayerNorm}(H_l \oplus \zeta H_{a-l}) 
\end{equation} 

\begin{equation} 
h^c_{l} = \text{AddNorm}(H^{c}_l \oplus \text{FeedForward}(H^{c}_l))
\end{equation} 
where \( \zeta H_{a-l} \) and \( \zeta H_{l-a} \) represent the propagated information from audio to text and from text to audio, respectively, both with dimensions of 1024. The variables \( h^c_{a} \) and \( h^c_{l} \) denote the enhanced cross-attention representations for audio and text, respectively, and their dimensions are also both 1024. $d_k$
represents the dimension of the key vector.

\subsubsection{Correlative Attention Network}
The correlative attention network is designed to enhance sentiment analysis by explicitly modeling the relationships between uni-modal and bimodal representations. It takes as input the latent uni-modal representations from the interactive correlation network and correlates them with the bimodal representation to capture both intra-modal and inter-modal dependencies. This network is crucial for bridging the gap between individual modality features and their combined representation, ensuring that the interactions between modalities are accurately captured.

The correlative attention network comprises two submodules: the joint attention network and the bimodal correlation evaluation network (see Fig. 5). The joint attention network integrates the uni-modal latent features from the audio and text modalities to produce a comprehensive bimodal representation, allowing the model to focus on the most salient features from each modality. Meanwhile, the bimodal correlation evaluation network assesses the correlations between these modalities by quantifying their interdependencies, offering a deeper understanding of how features from one modality influence those of the other.

By incorporating the correlative attention network, the model achieves a more robust and nuanced representation of the input data, improving its ability to analyze sentiment effectively. This approach ensures that the unique contributions of each modality are not only preserved but also synergistically leveraged to enhance overall performance.

\begin{figure}
    \centering
    \includegraphics[width=1\linewidth, height=0.35\textheight]{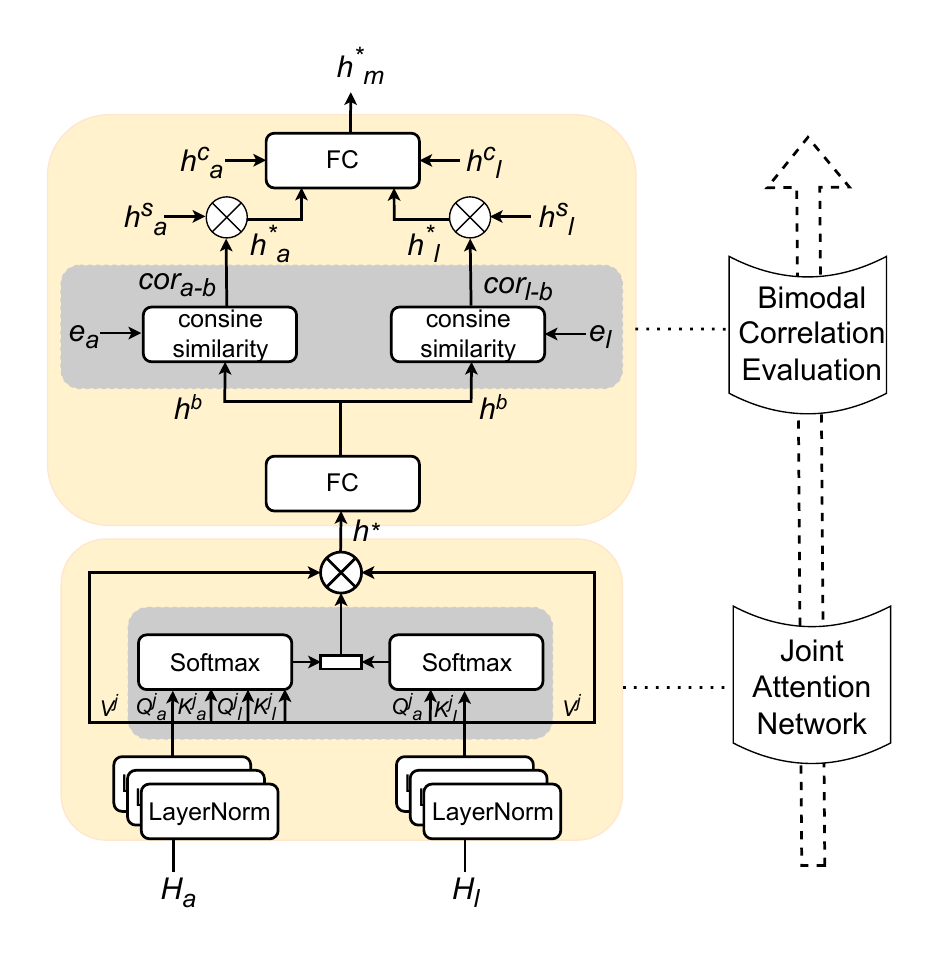}
    \caption{Update scheme of the correlative attention network consisting of the joint attention network and the bimodal correlation network.}
    \label{fig:correlation_attention_network}
\end{figure}

\textbf{Joint Attention Network}: The joint attention network aims to enhance the cross-modal relationship between audio and text by reducing noise and highlighting meaningful interactions. This is achieved using a pair of softmax functions, which help focus on the most relevant features from both modalities while suppressing less informative or noisy elements. The network effectively refines the cross-modal representation, enabling the model to better capture salient patterns for downstream tasks such as sentiment analysis.

The attention mechanism operates by mapping query, key, and value vectors to outputs. Specifically, it computes attention scores using query and key vectors and applies these scores to calculate a weighted sum of the value vectors. Given the uni-modal representations \( H_m \), the joint attention network first projects these representations into query, key, and value vectors (\( Q^j_a, Q^j_l, K^j_a, K^j_l, V^j_a, V^j_l \)) as follows:

\begin{equation}
Q^j_m = h^s_m W^{Q^j}_m, \quad K^j_m = h^s_m W^{K^j}_m, \quad V^j_m = h^s_m W^{V^j}_m
\end{equation} 

The joint attention network computes the joint bimodal representation using the following equation:
{\scriptsize
\begin{equation}
h^* = \left( \text{softmax}\left( \frac{Q^{j}_{a} K^{j \top}_{a} + Q^{j}_{l} K^{j \top}_{l}}{\sqrt{d_k}} \right) - \lambda \, \text{softmax}\left( \frac{Q^{j}_{a} K^{j \top}_{l}}{\sqrt{d_k}} \right) \right) V^{j}
\end{equation}}
where \( h^* \) denotes the joint bimodal representation with a dimension of 1024. \( \lambda \) is a learnable scalar that adjusts the weight of the second term, which represents cross-modal attention. \( V^j \) represents the average of the value vectors \( V^j_a \) and \( V^j_l \), encapsulating information from both modalities. $d_k$ represents the dimension of the key vector. 

The use of a pair of softmax functions serves a dual purpose. The first softmax term emphasizes intra-modal relationships by combining attention scores from the query and key vectors within each modality. This term ensures that strong intra-modal signals are preserved. The second softmax term focuses on cross-modal relationships, computing the attention scores between the query of one modality and the key of the other. By subtracting the cross-modal scores (weighted by \( \lambda \)), the network suppresses noisy or irrelevant interactions while retaining meaningful cross-modal dependencies. This design improves the robustness of the joint attention mechanism and enhances the overall representation quality for tasks requiring fine-grained cross-modal understanding.

\textbf{Bimodal Correlation Evaluation}: The latent uni-modal representations from the interactive correlation network (\(e_a\) for audio and \(e_l\) for text), along with the joint bimodal representation from the joint attention network (\(h^*\)), are input into the bimodal correlation evaluation network to assess the correlations between audio and text. Inspired by CLIP \cite{radford2021learning}, cosine similarity is employed to measure the correlations between different modalities, resulting in the pairwise cross-modal correlation coefficients, \( \text{cor}_{a-m} \) and \( \text{cor}_{l-m} \) (see Fig. 4). The joint multi-modal representation \( h^* \) is passed through a linear layer, reducing its dimension to 512, resulting in \( h^b \), which serves as the fused representation used for bimodal correlation evalution.

\begin{equation} 
\text{cor}_{a-b} = \left\langle e_a, h^b \right\rangle,  \quad \text{cor}_{l-b} = \left\langle e_l, h^b \right\rangle
\end{equation} 
where \( \left\langle \cdot \right\rangle \) denotes cosine similarity, and \( \text{cor}_{a-b} \) and \( \text{cor}_{l-b} \) represent the correlation coefficients for the audio-bimodal and text-bimodal pairs, respectively. 

The pairwise cross-modal correlation coefficients are subsequently integrated into the enhanced self-attention representations (\(h^s_a\) for audio and \(h^s_l\) for text) to produce correlative representations. This process ensures that the resulting representations fully capture inter-modal interactions by utilizing the correlation information to strengthen the interplay between modalities. Through this integration, the model becomes better equipped to align and fuse audio and text features effectively, thereby facilitating a more comprehensive understanding of cross-modal relationships, as described below:

\begin{equation} 
    h^*_a = h^s_a \times \text{cor}_{a-b},  \quad  h^*_l = h^s_l \times \text{cor}_{l-b}
\end{equation} 
where \( h^*_a \) and \( h^*_l \) denote the correlative uni-modal representations, enhanced by their respective modality correlation coefficients. These coefficients represent the weighted importance of one modality's features in the context of the other, effectively capturing the interdependence between modalities. The dimensions of both \( h^*_a \) and \( h^*_l \) are 512.

Finally, \( h^c_a \), \( h^c_l \), \( h^*_a \), and \( h^*_l \) are concatenated to form the aggregated bimodal representation, which is then passed through four fully connected layers for classification.

\begin{equation} 
    h^*_m =   h^c_a \; {\oplus} \; h^c_l \; {\oplus} \; h^*_a \; {\oplus} \; h^*_l
\end{equation} 
where \( h^*_m \) represents the aggregated bimodal representation with a dimension of 3072.

\subsection{Classification}
The overall optimization objective consists of the connection loss \( \mathcal{L}_c \), the audio loss \( \mathcal{L}_a \), the text loss \( \mathcal{L}_l \), and the bimodal loss \( \mathcal{L}_m \). Specifically, the audio loss, the text loss, and the bimodal loss are processed independently to generate predictions, which are then combined to make a final decision. 

\subsection{Training}
For the classification task, the audio loss, the text loss, and the bimodal loss are defined as the standard cross-entropy loss. Finally, the overall loss function is expressed as a linear combination of \( \mathcal{L}_c \), \( \mathcal{L}_a \), \( \mathcal{L}_l \), and \( \mathcal{L}_m \):

\begin{equation}
\mathcal{L} = \alpha \mathcal{L}_c + \beta (\mathcal{L}_a +  \mathcal{L}_l) + \mathcal{L}_m
\end{equation}
where \( \alpha \), and \( \beta \) are weighting factors that balance the contribution of each loss term.
\begin{itemize}
    \item \textbf{The audio loss} \( \mathcal{L}_a \) is used to capture distinctive audio information for emotion prediction: 

\begin{equation}
\hat{y}_a^c = \text{softmax}(W^a_e h^s_a + b^a_e) \quad 
\end{equation}

\begin{equation}
\mathcal{L}_a = - \sum_{c=1}^C y_a^c \log(\hat{y}_a^c) \quad 
\end{equation}

where the parameters $W^a_{e}$ and $b^a_{e}$ are learnable weights and biases. \( c \) represents the emotion categories, and \( \hat{y}_a^c \) and \( y_a^c \) denote the predicted and true labels, respectively.

\item \textbf{The text loss} \( \mathcal{L}_l \) is used to capture distinctive textual information for emotion prediction:

\begin{equation}
\hat{y}_l^c = \text{softmax}(W^l_e h^s_l + b^l_e)
\end{equation}

\begin{equation}
\mathcal{L}_{l} = - \sum_{c=1}^{C} y_l^c \log(\hat{y}_l^c)
\end{equation}
where the parameters $W^l_{e}$ and $b^l_{e}$ are learnable weights and biases. \( c \) represents the emotion categories, and \( \hat{y}_l^c \) and \( y_l^c \) denote the predicted and true labels, respectively.

\item \textbf{The bimodal loss} \( \mathcal{L}_m \) is used to capture interactions between modalities for emotion prediction:

\begin{equation}
\hat{y}_m^c = \text{softmax}(W^m_e h^*_m + b^m_e)
\end{equation}

\begin{equation}
\mathcal{L}_{m} = - \sum_{c=1}^{C} y_m^c \log(\hat{y}_m^c)
\end{equation}
where the parameters $W^m_{e}$ and $b^m_{e}$ are learnable weights and biases. \( c \) represents the emotion categories, and \( \hat{y}_m^c\) and \( y_m^c \) denote the predicted and true labels, respectively.

\end{itemize}
\section{Experiments}
\subsection{Database and Metrics}
We evaluated our proposed BCAF method on the two benchmark datasets: MELD \cite{poria2018meld} and IEMOCAP \cite{busso2008iemocap}. Both these two commonly used public datasets are multi-modal, containing audio, text and video modality for every utterance. Due to the natural imbalance across various emotions, we choose weighted average F1-score as the evaluation metric. Table I shows the distribution of train and validation and test samples for both two datasets. 

\begin{itemize}
\item MELD is a multi-modal and multi-party dataset for conversational emotion recognition \cite{poria2018meld}. It consists of 13,708 utterances in 1,433 dialogues collecting  from the Friends TV shows. Each utterance is labeled with one in seven emotions: anger, joy, sadness, neutral, disgust, fear and surprise. 

\item IEMOCAP contains videos of dyadic conversations of ten speakers, spanning 12.46 hours \cite{busso2008iemocap}. Each utterance is annotated using the following discrete categories: happy, sad, neutral, angry, excited, and frustrated. 
\end{itemize}

\begin{table}[t]
\centering
\footnotesize % Reduce the font size
\caption{Statistics of two benchmark datasets: MELD and IEMOCAP}
\label{tab:dataset_statistics}

\begin{tabular}{lcc|cc}
\toprule

\textbf{Dataset} & \multicolumn{2}{c}{\# Conversations} & \multicolumn{2}{c}{\# Utterances} \\
                 & \textbf{Train/Validation} & \textbf{Test} & \textbf{Train/Validation} & \textbf{Test} \\
\midrule
MELD~\cite{poria2018meld} & 1,153 & 280 & 11,098 & 2,610 \\
IEMOCAP~\cite{busso2008iemocap} & 120 & 31 & 5,810 & 1,623 \\
\bottomrule
\end{tabular}
\end{table}

% \begin{table}[t]
% \centering
% \footnotesize % Further reduce the font size
% \caption{Statistics of two benchmark datasets: MELD and IEMOCAP}

% \begin{tabular}{lcc|cc}
% \toprule
% \textbf{Dataset} & \multicolumn{2}{c}{\# Conversations} & \multicolumn{2}{c}{\# Utterances} \\
%                  & \textbf{train/validation} & \textbf{test} & \textbf{train/validation} & \textbf{test} \\
% \midrule
% MELD \cite{poria2018meld} & 1153 & 280 & 11098 & 2610 \\
% IEMOCAP \cite{busso2008iemocap} & 120 & 31 & 5810 & 1623 \\
% \bottomrule
% \end{tabular}
% \end{table}

%We evaluated the performance of multi-modal emotion recognition tasks using various metrics, including the weighted F1-score, defined as follows:

% \begin{equation}
% \textit{Precision} = \frac{\text{TP}}{\text{TP} + \text{FP}} 
% \end{equation}

% \begin{equation}
% \textit{Recall} = \frac{\text{TP}}{\text{TP} + \text{FN}} 
% \end{equation}

%\begin{equation}
%F1\text{-score} = 2 \times \frac{\text{Precision} \times \text{Recall}}{\text{Precision} + \text{Recall}} 
%\end{equation}

%The F1-score balances \textit{Precision} and \textit{Recall}, representing their harmonic mean. It is particularly useful when evaluating tasks with class imbalance, as it provides a more comprehensive assessment of the model's performance.

Given the inherent imbalance across different emotion classes, we used the weighted F1-score as our primary evaluation metric. The weighted F1-score calculates the F1-score for each class and applies weights based on the proportion of samples in each class. This approach ensures that the evaluation metric reflects the contribution of each class to the overall performance, addressing the impact of class imbalance effectively.

\subsection{Baseline and State-of-the-Art Methods} 

\begin{figure*}
\centering
\includegraphics[width=0.8\linewidth]{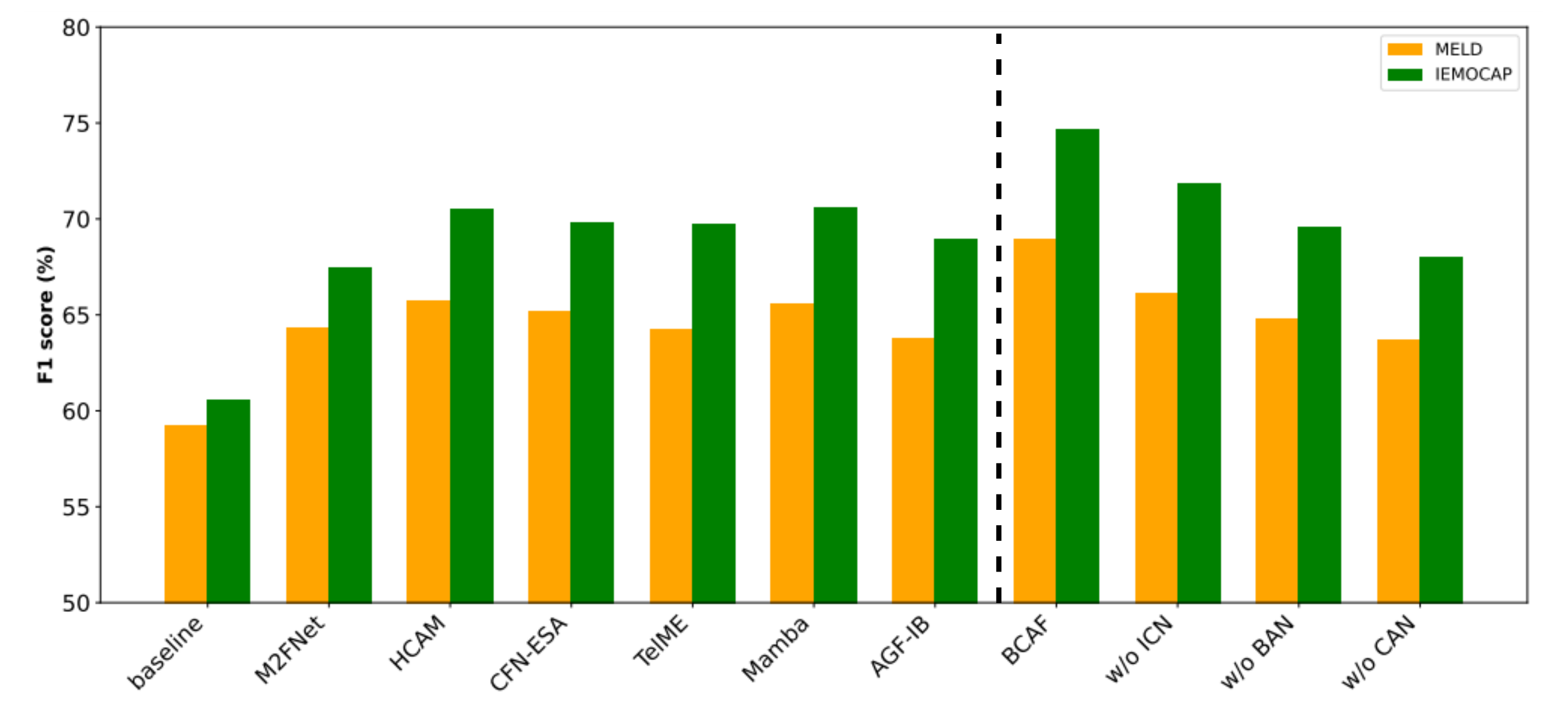}
\caption{Overall performance comparison between the BCAF method, the state-of-the-art methods, and the baselines.}
\end{figure*}

To comprehensively evaluate the performance of the proposed method, we compared our results with those of the bidirectional LSTM (bc-LSTM) baseline \cite{poria2018meld}. This baseline system leveraged an utterance-level LSTM to model context-aware representations from surrounding utterances. Additionally, we compared the proposed BCAF method to various existing state-of-the-art methods:

\begin{itemize}

\item \textbf{M2FNet} \cite{chudasama2022m2fnet} employed a multi-head attention-based fusion mechanism to combine emotion-rich latent representations of emotion-relevant features from visual, audio, and text modalities, learning both intra- and inter-modal relationships. 

\item \textbf{HCAM} \cite{dutta2023hcam} used a combination of recurrent and co-attention neural network to capture intra- and inter-modal interactions for emotion classification. 

\item \textbf{CFN-ESA} \cite{li2024cfn} incorporated a cross-modal fusion network with emotion-shift awareness for dialogue emotion recognition. 

\item \textbf{TelME} \cite{yun2024telme} used cross-modal knowledge transfer, using a language model (as the teacher) to enhance non-verbal modalities (as the 
student), thereby optimizing the performance of weaker modalities. 

\item \textbf{Mamba} \cite{shou2024revisiting} designed a multi-modal fusion strategy based on probability guidance to maximize information consistency across modalities and capture intra- and inter-modal interactions for conversational emotion recognition. 

\item \textbf{AGF-IB} \cite{shou2024adversarial} introduced graph contrastive representation learning to capture intra- and inter-modal complementary semantic information, as well as intra-class and inter-class boundary information for emotion categories.

\end{itemize}

\subsection{ Model Configuration}
We implemented our proposed BCAF method using the PyTorch 1.11.0 framework. The model was trained with the Adam optimizer with an initial learning rate of 1e-4 and an early-stopping strategy with a patience of 15 epochs. To aid convergence and improve generalization, we applied $L_2$ regularization with a weight of 0.0001 and used dropout with a rate of $p = 0.3$ to mitigate overfitting.

\section{Results and Discussion}
We start by conducting comparative experiments against previous state-of-the-art baselines. Next, we ablate core module to verify the effectiveness of our proposed BCAF method. Following that, we emphasize the importance of bimodal attention network and qualitative analysis. Finally, we conduct case studies and error analysis.

\subsection{Comparison with State-of-the-art Baselines}
We compared our proposed BCAF model against state-of-the-art baseline systems on two datasets, MELD and IEMOCAP. As shown in Fig. 6, BCAF demonstrates superior performance compared to state-of-the-art baseline systems in terms of the weighted F1-score, achieving a 3.15\% improvement over HCAM \cite{dutta2023hcam} on the MELD dataset and a 4.11\% improvement over Mamba \cite{shou2024revisiting} on the IEMOCAP dataset. These encouraging results demonstrate the superior expressive power and effectiveness of integrating the interactive connection network, bimodal attention network, and correlative attention network for bimodal speech emotion recognition.

Our BCAF model showed significant improvements over contextual models such as bc-LSTM \cite{poria2018meld}, M2FNet \cite{chudasama2022m2fnet}, and TelME \cite{yun2024telme}. We attribute this improvement to the fact that many contextual models fail to effectively learn modality connections and model intra- and inter-modality interactions between audio and text. In contrast, BCAF addresses these interactions through its interactive connection network, bimodal attention network, and correlative attention network. Specifically, BCAF demonstrated a strong ability to infer major emotion categories, such as \textit{neutral} and \textit{joy}, though it occasionally misclassified minority classes, such as \textit{anger}, on both the MELD and IEMOCAP datasets. We suggest that this may be due to the implicit expression and limited sample size of these emotions.

Our BCAF model exhibited smaller performance improvements on the MELD dataset compared to the IEMOCAP dataset. Specifically, while BCAF outperformed baseline systems on both datasets, the relative performance gain over state-of-the-art baseline systems (M2FNet \cite{chudasama2022m2fnet} on MELD versus TelME \cite{yun2024telme} on IEMOCAP) was more pronounced on the IEMOCAP dataset. Upon further analysis, we observed that dialogues in the MELD dataset are relatively shorter, typically consisting of 5 to 9 utterances, whereas dialogues in the IEMOCAP dataset average around 70 utterances per dialogue. Additionally, the MELD dataset, derived from real-world scenarios, contains significant background noise (e.g., honking, barking). Such noise, often uncontrollable and undesirable, can hinder the ability of individual modalities to effectively capture and convey emotional information. Furthermore, this noise may have introduced interference in the model, particularly for emotions like \textit{fear} and \textit{frustration}, which are more susceptible to being masked by background noise. As a result, the BCAF model achieved relatively better performance on the IEMOCAP dataset, where dialogues are longer and less affected by background noise, enabling the model to better capture emotional nuances.

\subsection{Ablation Study}
In this section, we conducted ablation studies to verify the effectiveness of our proposed BCAF model. We systematically ablated the model’s interactive connection network, bimodal attention network, and correlative attention network (see Fig. 6). Additionally, we investigated the contribution of each component by removing it from BCAF. Overall, we observed that the full version of BCAF achieved the best performance on both the MELD and IEMOCAP datasets. Particular emphasis was placed on the removal of the interactive connection network, the bimodal attention network, and the correlative attention network, each of which adversely impacted the model’s results. 

Specifically, the performance decline exhibits a consistent pattern when different components are removed from the BCAF model: the correlative attention network \textgreater{} the bimodal attention network \textgreater{} the interactive connection network (see Fig.~6). This suggests that the integration of these three core networks significantly enhances the performance of bimodal speech emotion recognition. The results demonstrate that our proposed BCAF method improves the representation capability of multi-modal features and effectively models correlations, as well as intra- and inter-modal interactions,between audio and text.

\subsubsection{Effect of the Interactive Connection Network} 
We first investigated the effect of the interactive connection network in our proposed BCAF. The interactive connection network was designed to leverage an encoder-decoder architecture to learn modality-specific features and capture modality connections between audio and text. 

We observed that the interactive connection network contributed to a performance improvement of 2.81\% on the MELD dataset and 2.83\% on the IEMOCAP dataset compared to the BCAF model without this component (see Fig. 6). This highlights the importance of the interactive connection network in enhancing the model's ability to capture and model intra- and inter-modal interactions between audio and text, leading to more accurate bimodal speech emotion recognition.

In the interactive connection network, each modality was processed using an encoder-decoder architecture to extract modality-specific features and analyze modality connections between audio and text. Specifically, the encoder transformed the raw input into a latent representation that captured essential characteristics specific to each modality. The decoder then reconstructed the original input from this latent space, effectively preserving intrinsic features within each modality, such as nuances in speech patterns or subtleties in textual expressions.

The connection loss encourages the BCAF model to learn modality connections by capturing shared features, enhancing emotion understanding. Previous approaches often processed modalities independently \cite{li2024cfn, shou2024adversarial}, potentially overlooking synergistic information from their interactions. By introducing regularization, the connection loss facilitates semantic complementation across modalities and maximizes the utility of modality connections.

\subsubsection{Effect of the Bimodal Attention Network} 
We then explored the role of the bimodal attention network, which aimed to leverage dynamic attention weights to learn intra- and inter-modal interactions between audio and text. Ablation results in Fig. 6 demonstrated an absolute improvement of 4.1\% on the MELD dataset and 5.12\% on the IEMOCAP dataset when the bimodal attention network was included in our BCAF model, compared to the version without this network. This improvement underscores the critical role of the bimodal attention network in enhancing the model's ability to focus on emotion-relevant features by effectively capturing interactions between audio and text. By selectively attending to important modality-specific information and their interactions, the bimodal attention network significantly contributed to improved bimodal speech emotion recognition performance.

In the ablation study of the multi-modal fusion module, we attributed the performance improvement to the bimodal attention network, which facilitated intra- and inter-modal interactions within and between audio and text while capturing long-term contextual information. By introducing dynamic self- and cross-modal attention weights to uni-modal representations, we enhanced these representations to build a robust and effective bimodal representation.

We proposed that the bimodal fusion module leveraged heterogeneous knowledge in a high-dimensional space to capture detailed information embedded in each modality while adaptively fusing implicit complementary content. This strengthened interactions and correlations, encouraging the model to explore complementary information and dynamic interactions between audio and text. As a result, the multi-modal fusion module significantly improved performance in bimodal speech emotion recognition tasks.

\subsubsection{Effect of the Correlative Attention Network} 
Finally, we examined the impact of the correlative attention network, which was designed to filter out noise in cross-modal relationships and learn inter-correlations between uni-modal and bimodal representations.

As shown in Fig.~6, removing the correlative attention network from the BCAF model resulted in a performance decrease of 5.24\% on the MELD dataset and 6.68\% on the IEMOCAP dataset compared to the complete model with all components included. This comparison underscores the significant role of the correlative attention network in improving emotion recognition by accurately integrating audio-text inter-correlations. The absence of this network hindered the model's ability to effectively filter noise and capture meaningful relationships across modalities.

The joint attention mechanism within the correlative attention network utilized a pair of softmax functions to simultaneously consider both self- and cross-attention weights for joint bimodal representation. This mechanism allowed the model to refine incorrect cross-modal information and dynamically adjust the influence of each modality, enabling comprehensive learning of intra- and inter-modal interactions between audio and text.
\begin{figure*}
\centering
\includegraphics[width=1\linewidth, height = 0.5\textheight]{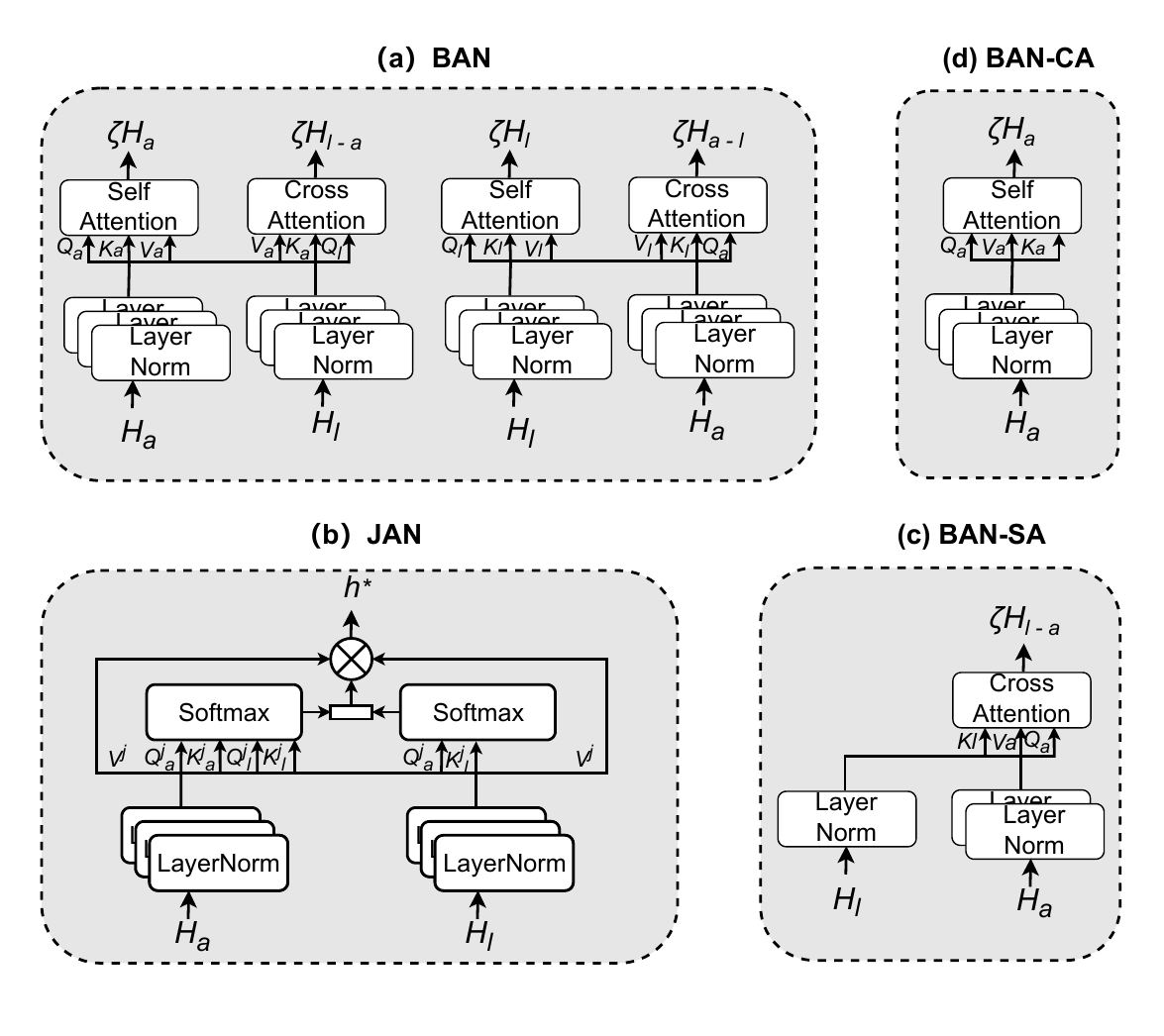}
\caption{Different attention variants: (a) BAN, (b) JAN, (c) BAN-SA, and (d) BAN-CA.}
\end{figure*}

\subsection{Impact of Attention}
% \begin{table*}[h!]
%     \caption{Impact of different attention mechanisms on our proposed BCAF model (Bold font indicates that BAN achieves the best F1 score).}
%     \label{tab:performance}
%     \centering
%     \resizebox{\textwidth}{!}{%
%     \begin{tabular}{lcccccc}
%         \toprule
%         \textbf{Attention} & \textbf{Equation} & \multicolumn{2}{c}{\textbf{w/o}} & \multicolumn{2}{c}{\textbf{Number of Layers}} \\
%         \cmidrule(lr){3-4} \cmidrule(lr){5-6}
%         & & \textbf{MELD} & \textbf{IEMOCAP} & \textbf{MELD} & \textbf{IEMOCAP} \\
%         \midrule
%         BAN & $\text{softmax}\left(\frac{Q_a K_a^\mathsf{T}}{\sqrt{d}}\right)V_a + 
%                 \text{softmax}\left(\frac{Q_a K_l^\mathsf{T}}{\sqrt{d}}\right)V_a + 
%                 \text{softmax}\left(\frac{Q_l K_l^\mathsf{T}}{\sqrt{d}}\right)V_l + 
%                 \text{softmax}\left(\frac{Q_a K_l^\mathsf{T}}{\sqrt{d}}\right)V_l$ & \textbf{64.83} & \textbf{69.57} & 5 & 3 \\
%         JAN & $\left( \text{softmax}\left( \frac{Q_a K_a^\top + Q_l K_l^\top}{\sqrt{d}} \right) - 
%                 \lambda \, \text{softmax}\left( \frac{Q_a K_l^\top}{\sqrt{d}} \right) \right) V$ & 66.08 & 71.35 & 3 & 4 \\
%         BAN-SA & $\text{softmax}\left(\frac{Q_a K_l^\mathsf{T}}{\sqrt{d}}\right)V_a$ & 67.21 & 72.16 & 7 & 5 \\
%         BAN-CA & $\text{softmax}\left(\frac{Q_a K_a^\mathsf{T}}{\sqrt{d}}\right)V_a$ & 67.52 & 73.01 & 7 & 6 \\
%         \bottomrule
%     \end{tabular}%
%     }

% \end{table*}
\begin{table*}[h!]
    \centering
    \caption{Impact of different attention mechanisms on our proposed BCAF model (Bold font indicates that BAN achieves the best F1 score).}
    \label{tab:performance}
    \resizebox{\textwidth}{!}{%
    \begin{tabular}{lcccccc}
        \toprule
        \textbf{Attention} & \textbf{Equation} & \multicolumn{2}{c}{\textbf{w/o}} & \multicolumn{2}{c}{\textbf{Number of Layers}} \\
        \cmidrule(lr){3-4} \cmidrule(lr){5-6}
        & & \textbf{MELD} & \textbf{IEMOCAP} & \textbf{MELD} & \textbf{IEMOCAP} \\
        \midrule
        BAN & $\text{softmax} \left( \frac{Q_a K_a^{\top}}{\sqrt{d}} \right) V_a + 
                \text{softmax} \left( \frac{Q_a K_l^{\top}}{\sqrt{d}} \right) V_a + 
                \text{softmax} \left( \frac{Q_l K_l^{\top}}{\sqrt{d}} \right) V_l + 
                \text{softmax} \left( \frac{Q_a K_l^{\top}}{\sqrt{d}} \right) V_l$ & \textbf{64.83} & \textbf{69.57} & 5 & 3 \\
        JAN & $\left( \text{softmax} \left( \frac{Q_a K_a^{\top} + Q_l K_l^{\top}}{\sqrt{d}} \right) - 
                \lambda \, \text{softmax} \left( \frac{Q_a K_l^{\top}}{\sqrt{d}} \right) \right) V$ & 66.08 & 71.35 & 3 & 4 \\
        BAN-SA & $\text{softmax} \left( \frac{Q_a K_l^{\top}}{\sqrt{d}} \right) V_a$ & 67.21 & 72.16 & 7 & 5 \\
        BAN-CA & $\text{softmax} \left( \frac{Q_a K_a^{\top}}{\sqrt{d}} \right) V_a$ & 67.52 & 73.01 & 7 & 6 \\
        \bottomrule
    \end{tabular}%
    }
\end{table*}

Furthermore, the bimodal correlation evaluation explicitly measured similarity by incorporating correlation coefficients into the uni-modal representations. This process aligned the uni-modal features with the joint bimodal space, enhancing the model's ability to capture and leverage complementary information across modalities for more accurate emotion recognition.

% \begin{table*}[h!]
%     \centering
%     \begin{tabular}{lcccccc}
%         \toprule
%         \textbf{Attention} & \textbf{Equation} & \multicolumn{2}{c}{\textbf{w/o}} & \multicolumn{2}{c}{\textbf{Number of Layers}} \\
%         \cmidrule(lr){3-4} \cmidrule(lr){5-6}
%         & & \textbf{MELD} & \textbf{IEMOCAP} & \textbf{MELD} & \textbf{IEMOCAP} \\
%         \midrule
%         BAN & $\text{softmax}\left(\frac{Q_a K_a^\mathsf{T}}{\sqrt{d}}\right)V_a + \text{softmax}\left(\frac{Q_a K_l^\mathsf{T}}{\sqrt{d}}\right)V_a + \text{softmax}\left(\frac{Q_l K_l^\mathsf{T}}{\sqrt{d}}\right)V_l + \text{softmax}\left(\frac{Q_a K_l^\mathsf{T}}{\sqrt{d}}\right)V_l$ & \textbf{64.83} & \textbf{69.57} & 5 & 3 \\
%         JAN & $\left( \text{softmax}\left( \frac{Q_a K_a^\top + Q_l K_l^\top}{\sqrt{d}} \right) - \lambda \, \text{softmax}\left( \frac{Q_a K_l^\top}{\sqrt{d}} \right) \right) V$ & 66.08 & 71.35 & 3 & 4 \\
%         BAN-SA & $\text{softmax}\left(\frac{Q_a K_l^\mathsf{T}}{\sqrt{d}}\right)V_a$ & 67.21 & 72.16 & 7 & 5 \\
%         BAN-CA & $\text{softmax}\left(\frac{Q_a K_a^\mathsf{T}}{\sqrt{d}}\right)V_a$ & 67.52 & 73.01 & 7 & 6 \\
%         \bottomrule
%     \end{tabular}
%     \caption{Impact of different attention on our proposed BCAF model (Bold font indicates that BAN achieves the best F1 score).}
%     \label{tab:performance}
% \end{table*}

In this section, we examined the impact of different attention variants (see Fig.~ 7). For each modality, self-attention and cross-attention mechanisms were employed to capture intra- and inter-modal interactions between audio and text. To evaluate the effectiveness of the proposed attention mechanism in modeling these interactions, we implemented four comparison systems. The experimental results are presented in Fig. 65 and Table II.

\begin{itemize} 
    \item \textbf{Bimodal Attention Network (BAN):} This network employed separate softmax functions to independently apply self-attention and cross-attention weights, enabling distinct modeling of intra- and inter-modal interactions (see Fig.~ 7 (a)).

    \item \textbf{Joint Attention Network (JAN):} This network utilized a combined pair of softmax functions to simultaneously apply both self-attention and cross-attention weights, allowing for integrated modeling of intra- and inter-modal interactions (see Fig.~ 7 (b)).

    \item \textbf{Bimodal Attention Network - Self-Attention (BAN-SA):} This variant used only self-attention, omitting the cross-attention mechanism to focus solely on intra-modal interaction modeling (see Fig.~ 7 (c)).

    \item \textbf{Bimodal Attention Network - Cross-Attention (BAN-CA):} This variant used only cross-attention, omitting the self-attention mechanism to concentrate exclusively on inter-modal interaction modeling (see Fig.~ 7 (d)).
\end{itemize}

We observed that the BAN achieved the best performance on both the MELD and IEMOCAP datasets. As shown in Fig. 6 and Table II, the inclusion of the BAN contributed to a performance improvement of 3.6\% on the MELD dataset and 6.6\% on the IEMOCAP dataset compared to the baseline model without BAN. This comparison highlights the effectiveness of BAN in leveraging both intra-modal and inter-modal interactions between audio and text.

Additionally, the decline in performance followed a consistent trend: BAN $\text{\textgreater}$ JAN $\text{\textgreater}$ BAN-SA $\text{\textgreater}$ BAN-CA. This means that removing certain attention mechanisms resulted in progressively worse performance. Specifically, the absence of JAN caused a performance decrease of 2.85\% on the MELD dataset and 3.34\% on the IEMOCAP dataset. Similarly, excluding BAN-SA led to a decrease of 1.72\% on MELD and 2.53\% on IEMOCAP, while removing BAN-CA resulted in the smallest decline of 1.41\% on the MELD and 1.69\% on the IEMOCAP. These results underscore the relative importance of each attention mechanism in modeling effective intra- and inter-modal interactions between audio and text. 

\begin{figure*}
\centering
\includegraphics[width=1\linewidth, height = 0.5\textheight]{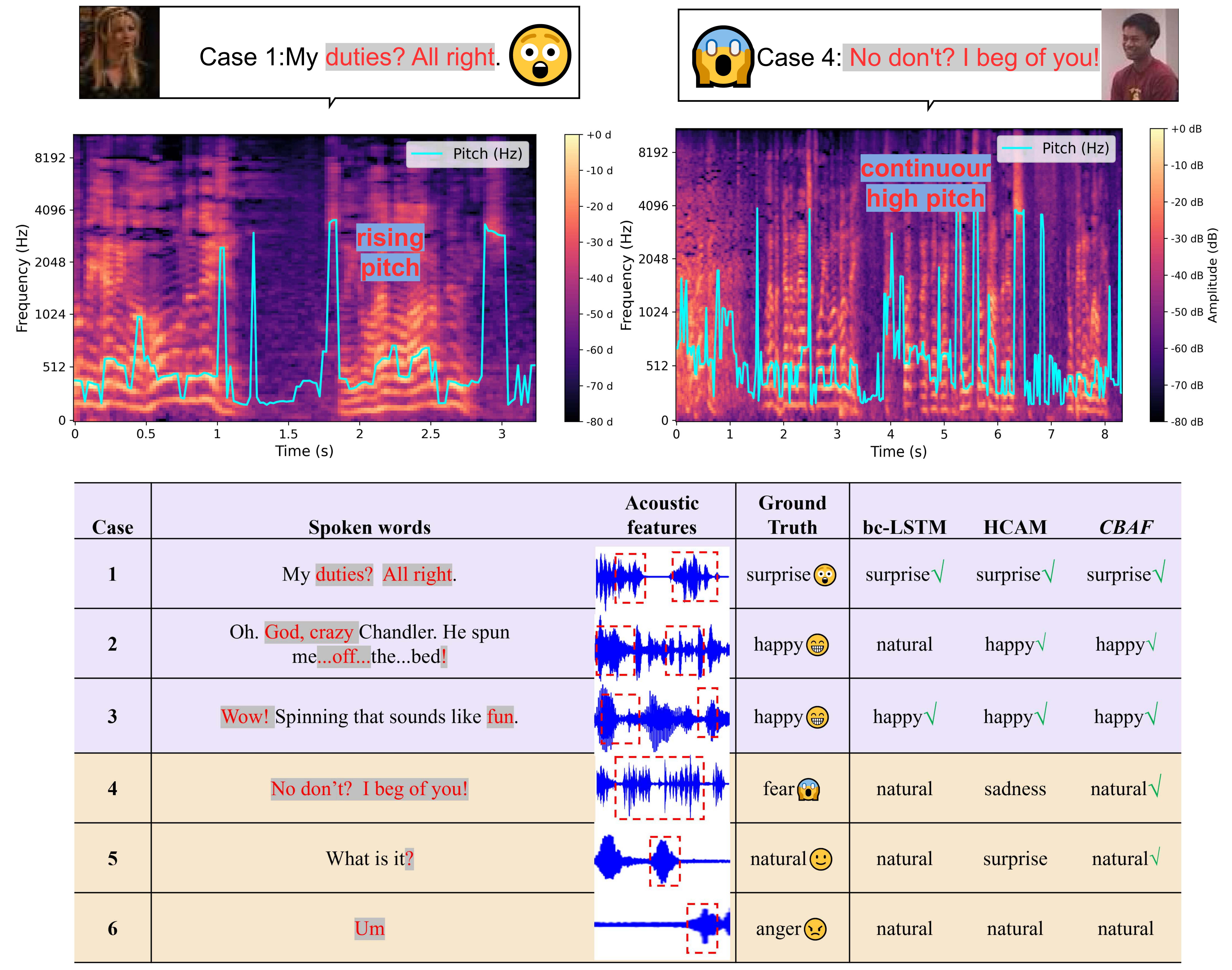}
\caption{Inputs from bimodal data and predictions using bc-LSTM, HACM, and our proposed BCAF method on the MELD and IEMOCAP datasets are analyzed in our case study.}
\end{figure*}

The bimodal attention network independently enabled complementary processing of intra-modal and inter-modal interactions between audio and text. We argue that intra-modal interactions provided the most essential information for accurate predictions, as they captured key modality-specific features directly tied to emotion recognition. Although inter-modal interactions added valuable complementary insights, core emotional cues were typically best represented within each individual modality, making intra-modal processing a primary factor in the model’s effectiveness. Moreover, maintaining sufficient layers in the BAN was crucial for optimal performance. Reducing the number of layers negatively impacted the model’s ability to process intra- and inter-modal interactions effectively, thereby degrading overall performance. 

% This further highlights the importance of balancing network depth to ensure both robustness and efficiency in bimodal emotion recognition tasks.

\subsection{Case Studies}
To illustrate the effectiveness of our proposed BCAF model, we selected samples from two datasets and conducted extensive experiments comparing BCAF with baseline models. The results, as shown in Fig. 8, revealed the following insights:

\begin{itemize}
    \item In most cases, our BCAF model achieved correct predictions, whereas the baseline models failed. This result indicates that BCAF effectively integrates audio information with the text modality, enhancing bimodal speech emotion recognition. By exploring correlations and intra- and inter-modal interactions between audio and text, BCAF provides richer, emotion-relevant insights, enabling the modalities to complement and augment each other.

    \item We observed challenges in predicting minor emotions when speech with significant background noise was introduced into the text modality for prediction. Specifically, ambiguous expressions, such as the single word \textit{“Um”}, were affected by unwanted and uncontrollable noise, further limiting the model's ability to comprehend human emotional expressions.

    \item Unexpectedly, while our proposed BCAF method outperformed the baseline models in terms of weighted F1-score, it performed worse than the state-of-the-art HCAM model. Upon further analysis, we attributed this to the emotional cues learned by the graph convolutional network in HCAM, which played a critical role in modeling and leveraging speaker information to capture intra- and inter-speaker dependencies for emotion recognition.
\end{itemize}

\begin{figure*}
\centering
\includegraphics[width=1\linewidth]{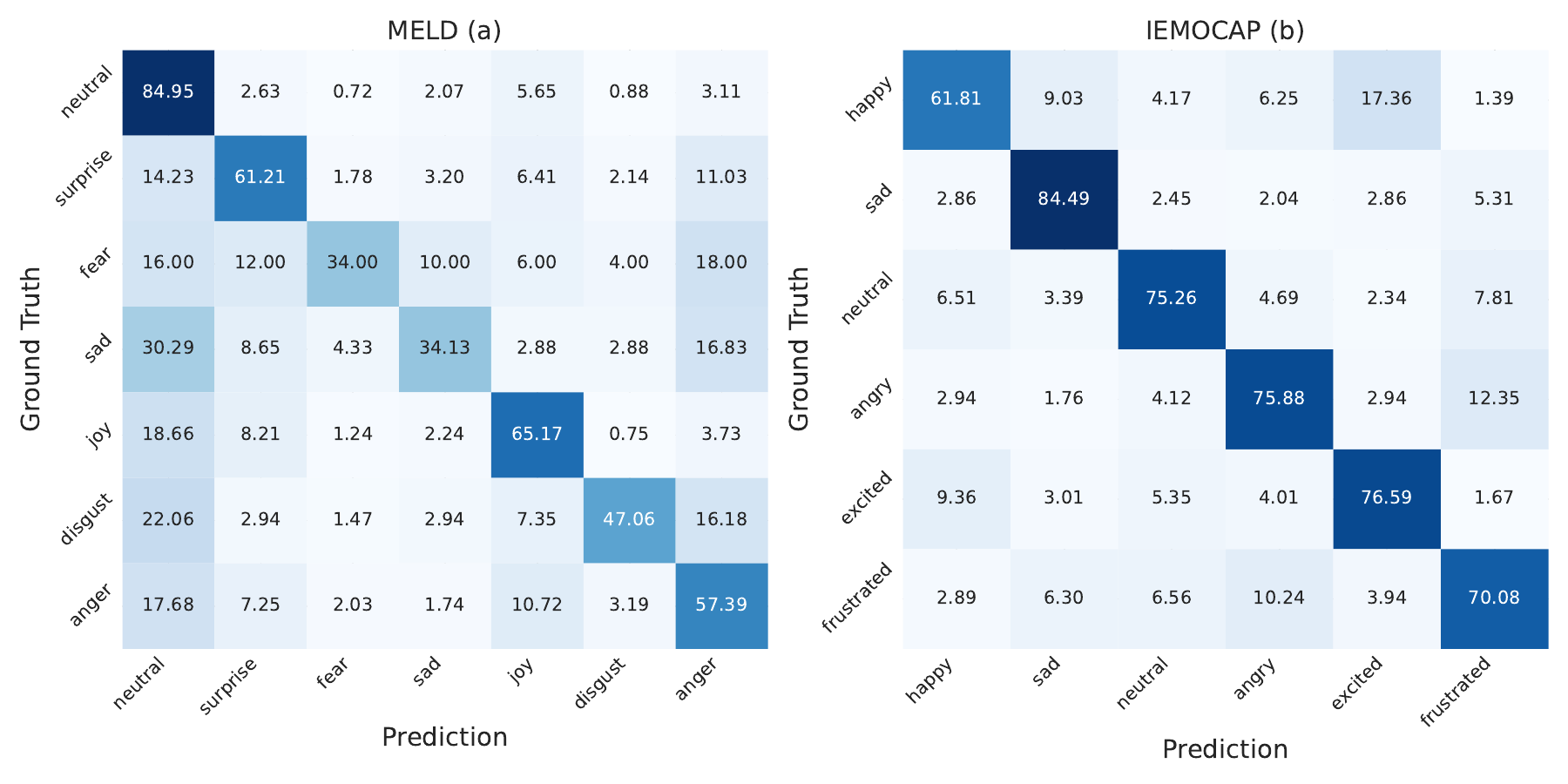}
\caption{Visualization of confusion matrices on the test sets of MELD (a) and IEMOCAP (b).}
\label{fig:res}
\end{figure*}

\subsection{Error Analysis}
We present the confusion matrices for the MELD and IEMOCAP datasets in Fig.~9. As shown, although our BCAF model achieved significant improvements over previous methods, as discussed in Section~V.A, its performance on several rare emotions (such as \textit{disgust}, \textit{fear}, and \textit{happiness}) remained unsatisfactory. Similar observations have been reported in previous studies, likely due to the implicit expression of these emotions.

By inferring predefined emotions in conversations, we observed that the errors made by the BCAF model were primarily caused by the following factors:

\begin{itemize}
    \item Emotions are complex interactions between subjective and objective factors, leading to potential biases in affect annotations. The emotion experienced by the speaker and the emotion perceived by human annotators may differ. Consequently, the BCAF model sometimes confused and misclassified similar or closely related emotions, such as \textit{fear} and \textit{sadness} (see Fig.~9).

    \item The MELD and IEMOCAP datasets closely mirror real-world scenarios, exhibiting significant class imbalances and substantial background noise. This observation aligns with the fact that humans remain neutral most of the time. In bimodal speech emotion recognition systems, this often resulted in minority emotions being predominantly misclassified as major classes.

    \item Humans express emotions through various modalities, including facial expressions, body posture, speech, and the linguistic content of verbal communication. However, the BCAF method relied solely on audio and text modalities for predicting emotions in conversations.
\end{itemize}

To address these challenges, we plan to implement several effective strategies, including resampling techniques, data augmentation, and transfer learning. Additionally, integrating information from multiple modalities in human communication could enhance the accuracy of emotion recognition, as different modalities complement and enrich each other, providing a more comprehensive set of emotion-relevant information.

\section{Conclusion}
We proposed the  Bimodal Connection Attention Fusion (BCAF) method for efficient and robust bimodal speech emotion recognition. BCAF consists of three key modules: the interactive connection network, the bimodal attention network, and the correlative attention network. The interactive connection network enables the model to learn modality connections between audio and text. The bimodal attention network facilitates semantic complementation and optimally leverages intra- and inter-modal interactions between audio and text. Finally, the correlative attention network filters noise from cross-modal relationships and learns inter-correlations between uni-modal and bimodal representations. Experimental results demonstrated that BCAF outperforms state-of-the-art methods in bimodal speech emotion recognition task.

% \section*{Acknowledgment}
% We would like to thank the anonymous reviewers for their valuable comments and feedback. Special acknowledgment is given to the Centre for Digital Music at Queen Mary University of London for its support. This research was funded by the China Scholarship Council and Queen Mary University of London.

\bibliographystyle{unsrt}  % or another style like plainnat, ieeetr, etc.

\bibliography{IEEEabrv,Bibliography}

% \bibliography{your_bib_file}  % Replace "your_bib_file" with the actual name of your .bib file (without .bib extension)

\vfill

\end{document}